\documentclass[twocolumn,aps,prl,superscriptaddress,english]{revtex4-2}

\usepackage[colorlinks=true,allcolors=blue]{hyperref}
\usepackage{array}
\usepackage{amsmath}
\usepackage{amssymb}
\usepackage{graphicx}
\usepackage{babel}
\usepackage{braket}
\usepackage{mathrsfs}
\usepackage{amsfonts}
\usepackage{epstopdf}
\usepackage{multirow}
\usepackage{color}
\usepackage{bm}
\usepackage{amsthm}

\usepackage{xcolor}
\usepackage{tikz,fp}
\usepackage{tikz-cd}
\usepackage{adjustbox}
\usepackage{enumitem}
\usepackage{makecell}
\usetikzlibrary{arrows}
\usetikzlibrary{intersections}
\usetikzlibrary{shapes.geometric}
\usetikzlibrary{decorations.pathmorphing, patterns,shapes}
\usetikzlibrary{decorations.markings}

\tikzset{
  mid arrow/.style={postaction={decorate,decoration={
        markings,
        mark=at position .575 with {\arrow[#1]{stealth}}
      }}},
  near arrow/.style={postaction={decorate,decoration={
        markings,
        mark=at position .275 with {\arrow[#1]{stealth}}
      }}},
   far arrow/.style={postaction={decorate,decoration={
        markings,
        mark=at position .800 with {\arrow[#1]{stealth}}
      }}},
}
\tikzset{snake it/.style={decorate, decoration=snake}}

\usetikzlibrary{arrows}
\usetikzlibrary{intersections}
\usetikzlibrary{shapes.geometric}
\usetikzlibrary{decorations.pathmorphing, patterns,shapes,fixedpointarithmetic}
\usetikzlibrary{decorations.markings}

\newcommand{\ee}{\mathrm{e}} %use for exponential
\newcommand{\ii}{{\rm i}} %use for imaginary number
\newcommand{\TT}{\mathcal{T}} %use for time reversal symmetry
\newcommand{\ZZ}{\mathbb{Z}} %use for integer
\newcommand{\UU}{\mathbb{U}} %use for symmetry on the whole system
\newcommand{\lrangle}[1]{\langle#1\rangle}
\newcommand{\abs}[1]{\left\lvert#1\right\rvert}
\newcommand{\tr}{\mathrm{tr}} 
\newcommand{\dg}{\dagger}
\newcommand{\vbra}[1]{\left(#1\right|}
\newcommand{\vket}[1]{\left|#1\right)}

\newcommand{\Varepsilon}{\mathcal{E}} %use for quantum operation
\newcommand{\KK}{\mathcal{K}} %use for Kaus operator
\renewcommand{\tilde}{\widetilde}
\renewcommand{\hat}{\widehat}

\renewcommand{\geq}{\geqslant}

\makeatletter
\def\maketitle{
\@author@finish
\title@column\titleblock@produce
\suppressfloats[t]
}
\makeatother
\begin{document}

\title{Detecting Quantum Anomalies in Open Systems}
\author{Yunlong Zang}
\affiliation{Kavli Institute for Theoretical Sciences, University of Chinese Academy of Sciences, Beijing 100190, China}
\author{Yingfei Gu}
\email{guyingfei@tsinghua.edu.cn}
\affiliation{Institute for Advanced Study, Tsinghua University, Beijing 100084, China}
\author{Shenghan Jiang}
\email{jiangsh@ucas.ac.cn}
\affiliation{Kavli Institute for Theoretical Sciences, University of Chinese Academy of Sciences, Beijing 100190, China}
\date{\today}

\begin{abstract}
Symmetries and quantum anomalies serve as powerful tools for constraining complicated quantum many-body systems, offering valuable insights into low-energy characteristics based on their ultraviolet structure.
Nevertheless, their applicability has traditionally been confined to closed quantum systems, rendering them largely unexplored for open quantum systems described by density matrices.
In this work, we introduce a novel and experimentally feasible approach to detect quantum anomalies in open systems.
Specifically, we claim that, when coupled with an external environment, the mixed 't Hooft anomaly between spin rotation symmetry and lattice translation symmetry gives distinctive characteristics for half-integer and integer spin chains in measurements of $\exp(\rm{i}\theta S^z_{\rm tot})$ as a function of $\theta$.
Notably, the half-integer spin chain manifests a topological phenomenon akin to the ``level crossing" observed in closed systems.
To substantiate our assertion, we develop a lattice-level spacetime rotation to analyze the aforementioned measurements. 
Based on the matrix product density operator and transfer matrix formalism, we analytically establish and numerically demonstrate the unavoidable singular behavior of $\exp(\rm{i}\theta S^z_{\rm tot})$ for half-integer spin chains. 
Conceptually, our work demonstrates a way to discuss notions like ``spectral flow'' and ``flux threading'' in open systems not necessarily with a Hamiltonian.  
\end{abstract}

\maketitle

\begin{table*}[t]
  \centering
  \begin{tabular}{| c | c | c| c | c|} 
  \hline
  & Symmetry condition & Symmetry twist & Characteristics & Spectral flow \\ 
  \hline
 ~Closed system~ & $\UU H\UU^\dg=H$ & ~Symmetry flux &  Energy spectrum & Energy level crossing\\ 
  \hline
  Open system & \makecell{$\mathbb{U}(g)\rho_0\mathbb{U}^\dg(g)=\rho_0$\\$\mathbb{U}(g)\,\KK_a\,\mathbb{U}^\dg(g)=\sum_b V_{ab}(g)\KK_b$} & ~Symmetry operator~ & ~Correlation spectrum~ & Singularity in $f(\theta)$ \\ 
  \hline
\end{tabular}
  \caption{A summary of the ``open-closed correspondence'' by a lattice level spacetime rotation (or modular transformation).  
      The symmetry of open system under consideration in this article is known as the ``weak'' symmetry, which commutes with the density matrix.
      The insertion of the symmetry operator, i.e. $\tr (\rho\,\ee^{\ii\theta S^z_{\rm tot}})$, corresponds to flux threading in closed system. The energy spectrum of the closed system corresponds to the spectrum of the transfer matrix which is named as ``correlation spectrum'' in this table, as it determines the correlation length of local operators.
      The occurrence of energy level crossings in the spectral flow of closed systems with quantum anomalies finds its counterpart in our open system anomaly, manifested as singularity of $f(\theta)$ defined in \eqref{eq:f_theta_def}.
  }
  \label{tab:1} 
\end{table*}

\emph{Introduction.} 
``Quantum anomaly''~\cite{Hooft1980} is a useful framework to organize robust low energy physics that is determined from their ultra-violet (UV) structures. 
For example, the fermion doubling theorem~\cite{nielsen1981} anticipates an even number of low-energy fermion modes when the system is defined on a lattice. In spin systems, the Lieb-Schultz-Mattis~(LSM) theorem~\cite{LiebSchultzMattis1961,Affleck1986,Oshikawa2000, Hastings2004} asserts that the ground states of a translational and spin rotational invariant half-integer spin chain must be gapless or spontaneous symmetry breaking.

In these examples, quantum anomaly is typically related to the spectrum of Hamiltonian.
However, it is crucial to recognize a quantum anomaly as an inherent characteristic of the Hilbert space and symmetry, transcending the specific selection of state or Hamiltonian. 
In this article, we will introduce a novel approach for identifying anomalies in open quantum systems~\cite{Viyuela2014two,Viyuela2014Uhlmann,HuangArovas2014topological,BudichDiehl2015topology,Bardyn2018probing,Altland2021symmetry,HuangSunDiehl2022topological,Ryu2024LSM,ZhouLiZhaiLiGu2023,Li2023numerical,Sieberer2023universality,HuangDiehl2024mixed}
drawing an analogy to the concept of ``level crossing'' in closed quantum systems~\cite{Oshikawa2000,Oshikawa2000topological,shiozaki2017matrix,shiozaki2018many,cian2021many,dehghani2021extraction}
\footnote{We note that there have been works discussing LSM theorem in open systems recently. Their approach is distinct from ours, which examines the degeneracy of stationary states of Lindblad equation~\cite{Ryu2024LSM}.}. 

\begin{figure}[t]
    \centering
    \includegraphics[scale=0.75]{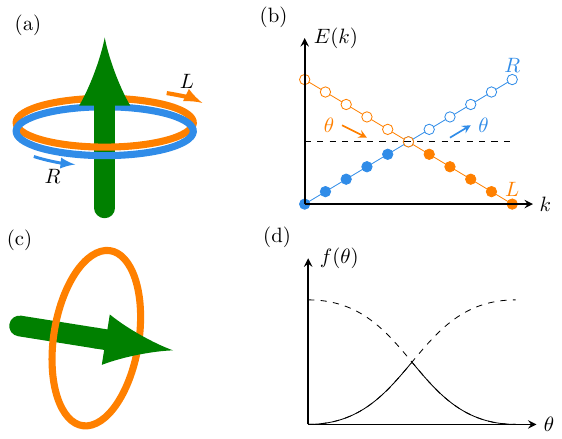}
    \caption{
    (a) A carton of two edges of an integer quantum Hall state with magnetic flux threaded. In real systems, the left-moving and right-moving edges are separated by the bulk of quantum Hall material, and are therefore protected from back-scattering.  
    (b) The corresponding single particle spectrum for the left and right moving modes under the flux threading.
    (c) Lattice level spacetime rotation maps the flux-threaded Hamiltonian of closed system to symmetry twisted transfer matrix. 
    (d) Schematic level crossing behavior of $f(\theta)$ as $\theta$ changes from 0 to $2\pi$, which corresponds to the crossing of the dominant eigenvalue of the transfer matrix under symmetry twist.}
    \label{fig:ftheta_schematic}
\end{figure}

A prominent manifestation of ``level crossing'' is observed in the edge of the integer quantum Hall state~\cite{Laughlin1981quantized,Wen1991}.
In this example, a $2\pi$ flux induces a spectral flow that raises or lowers a charge from or to the Fermi sea of two boundaries, and therefore leads to a unit charge pumping from one side to the other, as illustrated in Fig.~\ref{fig:ftheta_schematic}~(a,b). Consequently, the many-body energy spectrum will encounter a crossing when the single particle energy passes zero. 
The term ``anomaly'' in this context precisely characterizes nonzero charge pumping associated with the flux insertion~\cite{ Cheng2023}.

Now, a natural question arises when coupling the system to a bath, where energy is no longer conserved: Can we still discuss powerful tools such as ``spectral flow'', ``flux threading'' to detect quantum anomaly in the absence of ground states or Hamiltonians?
In the following, we propose an open-closed correspondence by a lattice-level spacetime rotation where the flux-threaded Hamiltonian of a closed system corresponds to the symmetry twisted spatial transfer matrix of a density matrix.
Furthermore, this innovative definition of spectral flow in terms of spatial transfer matrix will lead to prediction to new observables, see Fig.~\ref{fig:ftheta_schematic}(c) for an illustration and Table~\ref{tab:1} for a summary.
With this, we are able to detect quantum anomalies in open quantum systems in an experimentally feasible way.

\emph{Statement of results.}
    We examine a half-integer spin chain {\it symmetrically} coupled to an external environement.
    The dynamics of such system is described by density matrix $\rho(t)$.
    We claim that for a {\it short-range correlated} density matrix $\rho(t)$, function $f(\theta)$, defined as
    \begin{equation}
        f(\theta) := -\lim_{L\to\infty}\frac{1}{L} \ln \abs{ \lrangle{\ee^{\ii\theta S_{\rm tot}^z}}}
        \label{eq:f_theta_def}
    \end{equation}
    where $L$ denotes the system size, and $S^z_{\rm tot}=\sum_{j=1}^L S^z_j$, must exhibit singularity within the domain $\theta \in (0,2\pi)$~(see Fig.~\ref{fig:ftheta_schematic}(d) for an illustration). 

We wish to offer the following clarifications regarding the aforementioned statement:
\vspace{-7pt}
\begin{enumerate}
    \item A generic time evolution of this open spin chain is described by $\rho(t)=\Varepsilon_t(\rho_0)$, with $\rho_0$ the initial state and $\Varepsilon_t$ a quantum operation.
        $\Varepsilon_t$ can be further decomposed using Kraus operators: $\Varepsilon_t=\sum_a \KK_a \otimes\KK_a^\dg$.

        The full system, including the system-bath coupling, is {\it spin rotational} and {\it translational} symmetric.
        Upon integrating out the environment, the symmetry condition on $\rho_0$ and $\Varepsilon_t$ reads 
        \begin{equation}
        \begin{aligned}
            \mathbb{U}(g)\rho_0\mathbb{U}^\dg(g)&=\rho_0 ~,\\
            \mathbb{U}(g)\,\KK_a\,\mathbb{U}^\dg(g)&=\sum_b V_{ab}(g)\KK_b
            \label{}
        \end{aligned}
        \end{equation}
        where $\mathbb{U}(g)$ represents symmetry action on the spin chain, and $V_{ab}(g)$ some gauge transformation.
    Consequently, $\rho(t)$ satisfies the ``weak symmetry'' condition \cite{buca2012note}
        \begin{equation}
            \mathbb{U}(g)\,\rho(t)\,\mathbb{U}^\dg(g) = \rho(t)~,
            \label{eq:weak_sym}
        \end{equation}
    \item The short-range correlation refers to the condition $\lrangle{\hat{O}_j\hat{O}_k}-\lrangle{\hat{O}_j}\lrangle{\hat{O}_k}\sim \ee^{-\abs{j-k}/\xi}$ for any local operators $\hat{O}_{j}$ and $\hat{O}_k$ acting on site $j$ and $k$ respectively. 
        Here $\xi$ is the correlation length which may arise from e.g. thermalization $\xi \sim \beta$ or gap in the full system $\xi \sim (\text{gap})^{-1}$.  
    \item The spin rotational and lattice translational symmetry, together with the Hilbert space structure of a half-integer spin chain give a mixed 't Hooft anomaly between these two symmetries~(also known as the LSM anomaly)~\cite{Furuya2017,Cho2017,Metlitski2018,Else2020,Cheng2023}, forbidding a symmetric gapped phase in the long-wavelength for closed system. 
        In fact, the subgroup $U(1) \rtimes \ZZ_2 $ (generated by $\ee^{i \theta S^z_{\rm tot}}$ and $\ee^{i \pi S^x_{\rm tot}}$) of the full spin rotation is sufficient for the anomaly discussion. 
    \item $f(\theta)\geq f(0)=f(2\pi)=0$ due to the unitarity of $\ee^{\ii \theta S^z_{\rm tot}}$.
        $f(\theta)$ can be a smooth function (no robust singularity) for an integer spin open chains~\cite{Sm}.
        Therefore, $f(\theta)$ provides a practical approach to distinguish between integer and half-integer spin chains in open systems. 
        It also detects the ``mismatch'' of strong and weak symmetry, as the former (i.e. $U\rho = e^{i\phi} \rho$) implies a constant $f(\theta) =1$~\footnote{On the other hand, a strong symmetry invariant density matrix can not be short-range correlated from the LSM theorem.}.
\end{enumerate}

\emph{Intuitions from modular invariant field theories.}
Here we present an intuition from modular invariant field theories~\cite{cardy1986186}.
For example, let us consider a thermal density matrix $\rho=Z^{-1} \ee^{-\beta H}$ of a $(1+1)$D conformal field theory with $U(1)$ symmetry whose charge is denoted as $Q$. 
We put the theory on a circle of length $L\gg \beta$, and therefore the partition function lives on a torus. 

\begin{figure}[t]
    \centering
    \includegraphics[width=0.9\columnwidth=1]{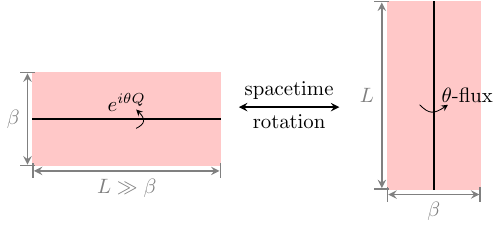}
    \caption{
        Space-time rotation or modular transformation in a context of a $(1+1)$D conformal field theory.
        The left panel illustrates a path integral representation of $\tr (\rho \ee^{\ii\theta Q})$ where the symmetry operator is applied on the Cauchy surface drawn as a black line.
        After spacetime rotation or modular transformation with space and time exchanged, the symmetry operator is mapped to a symmetry flux, and now the inverse temperature is $\tilde{\beta}=L\gg \beta$. 
        Therefore, the right hand side describes a flux-threaded Hamiltonian at very low temperature.
     As a result, the function $f(\theta)$ defined by the left side can be estimeated by $\ee^{-L f(\theta)} =\abs{ \tr(\rho \ee^{\ii\theta Q}) }\approx\ee^{-\tilde{\beta}E_\theta}$, so 
    $f(\theta)\approx E_\theta$.
        \iffalse
        As a result, the function $f(\theta)$ that is defined by the left side is related to the ground state energy $E(\theta)$ for systems with finite length.  
        Now, with the symmetry operator inserted, i.e. $\tr(\rho \ee^{\ii\theta Q})$, the torus is ``twisted'' in the temporal direction, as shown in Fig.
More explicitly, a charged operator will acquires $\theta$ phase after passing the line (Cauchy surface) where the symmetry operator acts on.
For a modular invariant theory, such picture can be rotated with space and time exchanged, depicted on the right hand side of Fig.
Then, the symmetry operator becomes a ``flux'' line in Hamiltonian, on a circle of length $\beta \ll L$ where $\tilde{\beta}=L$ is the new inverse temperature in the rotated picture.\fi
    }
    \label{fig:modular_transf_theta}
\end{figure}

Now, with the symmetry operator inserted, i.e. $\tr(\rho \ee^{\ii\theta Q})$, the torus is ``twisted'' in the temporal direction, which becomes spatial twisted boundary condition after rotation, as shown in Fig.~\ref{fig:modular_transf_theta}.
Therefore, we can estimate the function $f(\theta)$ as
$f(\theta)\approx E_\theta$\footnote{In general, $ \ee^{-L f(\theta)} \approx d_g\ee^{-\tilde{\beta}E_\theta}$, where $d_g$ is the degeneracy factor. This type trick, relating high-temperature properties to low-temperature, is known as Cardy formula. },
which becomes the ground state energy under twisted boundary condition when $L\to\infty$.
In Supplemental Material~\cite{Sm}, we review how quantum anomaly leads to non-analytical $E_\theta$. 
Then, a cusp naturally emerges as a consequence of the spectral flow of the original theory with anomaly(see Supplemental Material~\cite{Sm} for an in-depth explanation of the spectral flow of low-energy levels in an LSM system.).
\iffalse
Note that the ground state energy is a periodic function of $\theta$ due to charge quantization.
Then, a cusp naturally emerges as a consequence of the spectral flow of the original theory with anomaly, e.g. a mixed 't Hooft anomaly between the vector and axial ${\rm U}(1)$ symmetry, also known as chiral or ABJ anomaly~\cite{Adler1969,bellJackiw1969}.
In the context of the half-integer spin chain, $S^z_{\rm tot}$ plays a role analogous to that of the vector $U(1)$ charge $Q$ mentioned earlier, while the many-body momentum bears resemblance to the axial $U(1)$ charge~(see Supplemental Material~\cite{Sm} for an in-depth explanation of the spectral flow of low-energy levels in an LSM system.).
\fi

\emph{Lattice level modular transformation}.
However, the validity of the above argument crucially relies on the ability to interchange roles of ``space" and ``time", a condition not always available for generic many-body systems on the lattice.
In the following, we tackle this challenge with an approach that enables us to discuss ``flux threading'' and ``spectral flow'' for systems without modular invariance.
The key idea is to utilize the ``transfer matrix'' as a lattice version of the spacetime rotated, or modular transformed, Hamiltonian.
We summarize the correspondence in Table~\ref{tab:1}. 

Specifically, we construct lattice level modular transformation by representing the short-range correlated density matrix $\rho$ as a finite bond dimensional matrix product density operator~(MPDO)~\cite{Verstraete2004,Perez2007,Verstraete2006matrix,Cirac2017,AlhambraCirac2021,Huang2021,Cirac2021matrix,Chen2020matrix,Svetlichnyy2022a,Svetlichnyy2022b}:

\begin{equation}
\begin{aligned}
    \rho&=\sum_{\{s\},\{s'\}}\tr\left[ \cdots \hat{M}^{s_{j}s'_{j}}\hat{M}^{s_{j+1}s'_{j+1}}\cdots \right]\\
    & \qquad \qquad \times\ket{\dots s_j s_{j+1},\dots}\bra{\dots s'_j s'_{j+1} \dots}\\
    & = \qquad  
    \adjincludegraphics[scale=1,valign=c]{ 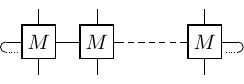}
    \iffalse\begin{tikzpicture}[scale=1,baseline={([yshift=-2pt]current bounding box.center)}]
    \draw[white] (0pt,-20pt) -- (0pt,20pt);
    \draw (-8pt,0pt) -- (46pt,0pt);
    \draw[densely dashed] (46pt,0pt) -- (78pt,0pt);
    \draw (78pt,0pt) -- (104pt,0pt);
    \draw (-8pt,0pt) arc (90:270:2.5pt);
    \draw (104pt,0pt) arc (90:-90:2.5pt);
    \draw [densely dotted] (-8pt,-5pt) -- (-2pt,-5pt);
    \draw [densely dotted] (104pt,-5pt) -- (98pt,-5pt);
    %\draw [opacity=0.5, densely dashed] (-8pt,-5pt) -- (104pt,-5pt);
    \draw (8pt,-16pt) -- (8pt,16pt);
    \draw (36pt,-16pt) -- (36pt,16pt);
    \draw (88pt,-16pt) -- (88pt,16pt);
    \filldraw[thick,fill=white] (0pt,-8pt) rectangle (16pt,8pt);
    \node at (8pt,0pt) {$M$};
    \filldraw[thick,fill=white] (28pt,-8pt) rectangle (44pt,8pt);
    \node at (36pt,0pt) {$M$};
    \filldraw[thick,fill=white] (80pt,-8pt) rectangle (96pt,8pt);
    \node at (88pt,0pt) {$M$};
    \end{tikzpicture}
    \fi
    \label{}
\end{aligned}
\end{equation}
where $\hat{M}^{ss'}=M^{ss'}_{\alpha\beta}\vket{\alpha}\vbra{\beta}$, with $\vket{\cdot}/\vbra{\cdot}$ denoting states in left/right virtual legs of the local tensor.
Note that all local tensors $M$ in the MPDO are identical due to translational symmetry. 

With the MPDO representation, we can define the spatial {\it transfer matrix} by contracting physical legs of $M$:
\begin{equation}
    \hat{T} \equiv \sum_{s} \hat{M}^{ss}
    = 
    \adjincludegraphics[scale=1,valign=c]{ 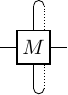}.
    \iffalse\begin{tikzpicture}[scale=1,baseline={([yshift=-2pt]current bounding box.center)}]
    \draw (8pt,-20pt) -- (8pt,20pt);
    \draw (-8pt,0pt) -- (24pt,0pt);
    \draw (8pt,20pt) arc (180:0:2.5pt);
    \draw (8pt,-20pt) arc (180:360:2.5pt);
    \draw [densely dotted] (13pt,20pt) -- (13pt,8pt);
    \draw [densely dotted] (13pt,-20pt) -- (13pt,-8pt);
    \filldraw[thick,fill=white] (0pt,-8pt) rectangle (16pt,8pt);
    \node at (8pt,0pt) {$M$};
    \end{tikzpicture}
    \fi
\end{equation}
The eigenvalues of  $\hat{T}$ are generally complex numbers and will be called \emph{correlation spectrum}. 
The modulus of these eigenvalues provides information about correlation lengths, while their arguments characterize oscillation wavevectors of correlators. 

\emph{Spectral flow on spatial transfer operator.}
Equipped with the MPDO representation, we have
\begin{equation}
    \begin{aligned}
    \lrangle{\ee^{\ii\theta S^z_{\rm tot}}}
    &=
    \adjincludegraphics[scale=1,valign=c]{ 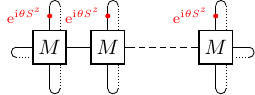}
\\
    &=:
    \tr\big[ \big( \hat{T}(\theta) \big)^L \big] 
    \end{aligned}
\end{equation}
where $L$ is the length of the system, i.e. the number of local tensor $M$, and the symmetry twisted transfer matrix $\hat{T}(\theta)$ is defined as 
\begin{equation}
   \hat{T}(\theta) 
   :=
   \adjincludegraphics[scale=1,valign=c]{ 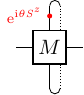}
   \iffalse
   \begin{tikzpicture}[scale=1,baseline={([yshift=-2pt]current bounding box.center)}]
    \draw (8pt,-20pt) -- (8pt,20pt);
    \draw (-8pt,0pt) -- (24pt,0pt);
    \draw (8pt,20pt) arc (180:0:2.5pt);
    \draw (8pt,-20pt) arc (180:360:2.5pt);
    \draw [densely dotted] (13pt,20pt) -- (13pt,8pt);
    \draw [densely dotted] (13pt,-20pt) -- (13pt,-8pt);
    \filldraw[thick,fill=white] (0pt,-8pt) rectangle (16pt,8pt);
    \node at (8pt,0pt) {$M$};
    \filldraw [red,fill=red] (8pt,15pt) circle (1pt) node[left] {\scriptsize $\ee^{i \theta S^z}$};
    \end{tikzpicture} \fi~~=\sum_{ss'}\left[ \exp(\ii \theta S^z) \right]_{s's}\cdot \hat{M}^{ss'}.
    \label{eq:Ttheta_def}
\end{equation}
As demonstrated in the Supplemental Material\cite{Sm} , the dominant eigenvalue in the correlation spectrum, defined by the eigenvalue of the untwisted transfer matrix $\hat{T}$ with the largest modulus, is non-degenerate for short-range correlated $\rho$, and equals $1$ due to the normalization condition $\tr(\rho)=1$. Now, we analogously define $\lambda_{\max}(\theta)$ as the dominant eigenvalue for $\hat{T}(\theta)$, then we have
\begin{equation}
     f(\theta)=-\ln\abs{\lambda_{\max}(\theta)}
\end{equation}
as $L\rightarrow \infty$. 
%Note that $f(\theta) \geq 0$ due to the unitarity of the symmetry representation. 

We proceed to examine the properties of $\lambda_{\max}(\theta)$ stemming from symmetries of $\hat{T}(\theta)$.
It is worth noting that the onsite symmetry $g$ are implemented as gauge transformations in virtual legs~\cite{Perez2008string}, and the local tensor $\hat{M}^{ss'}$ remains invariant under symmetry actions on all physical legs~(labeled as $U(g)$) and virtual legs~(labeled as $W(g)$), namely, 
\begin{equation}
    M^{ss'}_{\alpha\beta}=[U(g)]_{st} [U^*(g)]_{s't'} [W(g)]_{\alpha\gamma} [W^*(g)]_{\beta\delta} M^{tt'}_{\gamma\delta} 
\label{eq:local_tensor_sym}
\end{equation}
or pictorially 
\begin{equation}
\adjincludegraphics[scale=1,valign=c]{ 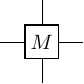}
    =
    \adjincludegraphics[scale=1,valign=c]{ 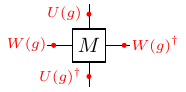}.
\iffalse
    \begin{tikzpicture}[scale=1,baseline={([yshift=-2pt]current bounding box.center)}]
    \draw (8pt,-20pt) -- (8pt,20pt);
    \draw (-12pt,0pt) -- (28pt,0pt);
    \filldraw[thick,fill=white] (0pt,-8pt) rectangle (16pt,8pt);
    \node at (8pt,0pt) {$M$};
    \end{tikzpicture}
    \fi
    \iffalse\begin{tikzpicture}[scale=1,baseline={([yshift=-2pt]current bounding box.center)}]
    \draw (8pt,-20pt) -- (8pt,20pt);
    \draw (-12pt,0pt) -- (28pt,0pt);
    \filldraw[thick,fill=white] (0pt,-8pt) rectangle (16pt,8pt);
    \node at (8pt,0pt) {$M$};
    \filldraw [red,fill=red] (8pt,15pt) circle (1pt) node[left] {\scriptsize $U(g)$};
    \filldraw [red,fill=red] (8pt,-15pt) circle (1pt) node[left] {\scriptsize $[U(g)]^\dagger$};
    \filldraw [red,fill=red] (-9pt,0pt) circle (1pt) node[left] {\scriptsize $W(g)$};
    \filldraw [red,fill=red] (25pt,0pt) circle (1pt) node[right] {\scriptsize $[W(g)]^\dagger$};
    \end{tikzpicture} .\fi
\end{equation}
From this equation, as well as $\ee^{\ii\pi S^x} \ee^{\ii \theta S^z}\ee^{-\ii\pi S^x}= \ee^{-\ii \theta S^z}$, we have 
\begin{equation}
\adjincludegraphics[scale=1,valign=c]{ 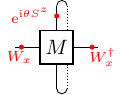}
    =
    \adjincludegraphics[scale=1,valign=c]{ 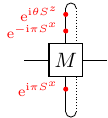}~
    =\adjincludegraphics[scale=1,valign=c]{ 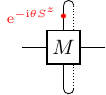},
\end{equation}
where $W_x$ labels $\pi$-rotation along $S^x$ axis acting on virtual legs. 

Hermiticity of $\rho$ establishes a relation between $\hat{M}^{ss'}$ and $\left[\hat{M}^{s's}\right]^*$ through some gauge transformation denoted as $J$\footnote{We here assume that the equality of two matrix product operators means their local tensors are related by gauge transformation.}:
\begin{equation}
\adjincludegraphics[scale=1,valign=c]{ 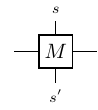}~~=
\adjincludegraphics[scale=1,valign=c]{ 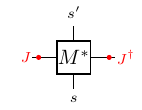}.
\iffalse
    \begin{tikzpicture}[scale=1,baseline={([yshift=-2pt]current bounding box.center)}]
    \draw (8pt,-15pt) -- (8pt,15pt);
    \draw (-12pt,0pt) -- (28pt,0pt);
    \filldraw[thick,fill=white] (0pt,-8pt) rectangle (16pt,8pt);
    \node at (8pt,0pt) {$M$};
    \node[above] at (8pt,15pt) {\scriptsize $s$};
    \node[below] at (8pt,-15pt) {\scriptsize $s'$};
    \end{tikzpicture}
    =
    \begin{tikzpicture}[scale=1,baseline={([yshift=-2pt]current bounding box.center)}]
    \draw (8pt,-15pt) -- (8pt,15pt);
    \draw (-12pt,0pt) -- (28pt,0pt);
    \filldraw[thick,fill=white] (0pt,-8pt) rectangle (16pt,8pt);
    \node at (8pt,0pt) {$M^*$};
    \node[above] at (8pt,15pt) {\scriptsize $s'$};
    \node[below] at (8pt,-15pt) {\scriptsize $s$};
    \filldraw [red,fill=red] (-9pt,0pt) circle (1pt) node[left] {\scriptsize $J$};
    \filldraw [red,fill=red] (25pt,0pt) circle (1pt) node[right] {\scriptsize $J^\dagger$};
    \end{tikzpicture} .\fi
\end{equation}
Therefore, together with Eq.~\eqref{eq:Ttheta_def}, we conclude
\begin{equation}
    \hat{T}(-\theta)=J\hat{T}^*(\theta)J^\dg
\end{equation}

To summarize, we obtain the following key relation 
\begin{equation}
    W_x \hat{T}(\theta)W^\dg_x 
    =J\hat{T}^*(\theta)J^\dg
    =\hat{T}(-\theta)=-\hat{T}(2\pi-\theta)
\label{eq:transfer_mat_z2}
\end{equation}
While the first equation holds true for both integer and half-integer spin systems, the minus sign in the last equation stems from the $2\pi$ rotation of the physical half-integer spin, serving as the central ingredient for the anomaly detection in half-integer spin chains.
The first equation suggests that eigenvalues of $\hat{T}(\theta)$, denoted as $\{\lambda_j(\theta)\}$, must either be real or come in conjugate pairs, while the minus sign implies that $\{\lambda_j(2\pi-\theta)\}$ and $\{\lambda_j(\theta)\}$ possess identical magnitudes but opposite signs.

In the following, we aim to demonstrate the inevitable singular behavior of $\lambda_{\rm max}(\theta)$ through a proof by contradiction. 
To initiate this demonstration, we commence tracing the spectral flow originating from $\lambda_{\rm max}(0)$, denoted as $\lambda_0(\theta)$, which is a smooth function with respect to $\theta$.
However, It is crucial to note that $\lambda_0(\theta)$ may be diverge from $\lambda_{\max}(\theta)$ due to level crossing, resulting in cusps in $f(\theta)$ at crossing points.

Our demonstration proceeds by assuming that $\lambda_0(\theta)$ remains non-degenerate, and equals $\lambda_{\max}(\theta)$ for all $\theta$.
In particular, $\lambda_0(\theta)$ ends at $\lambda_{\max}(2\pi)=-1$.
Flow of $\lambda_0$ is schematically depicted in the complex as follows:
\begin{equation*}
    ~\quad\adjincludegraphics[scale=1,valign=c]{ 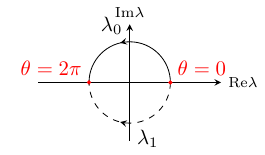}
    \iffalse
    \begin{tikzpicture}[scale=0.8]
        \draw[->,>=stealth] (-55pt,0pt) -- (55pt,0pt)node[right] {\scriptsize $\rm{Re}\lambda$};
        \draw[->,>=stealth] (0pt,-35pt) -- (0pt,35pt)node[above] {\scriptsize $\rm{Im}\lambda$}; 
        \draw[mid arrow](24.5pt,0) arc (0:180: 24.5pt and 24.5pt);
        \draw[mid arrow,dashed](24.5pt,0) arc (0:-180: 24.5pt and 24.5pt);
        \filldraw [red,fill=red] (24.5pt,0) circle (1pt) node[above right] {$\theta=0$};
        \filldraw [red,fill=red] (-24.5pt,0) circle (1pt) node[above left] {$\theta=2\pi$};
        \node[above left] at (0,24.5pt) {$\lambda_0$};
        \node[below right] at (0,-24.5pt) {$\lambda_1$};
    \end{tikzpicture}\fi
\end{equation*}
Note that $\lambda_0(\theta)$ exhibits relection symmetry under the imaginary axis, a consequence of the Hermiticity condition in Eq.~\eqref{eq:transfer_mat_z2}.
However, it is crucial to recognize the emergence of a second flow, denoted as $\lambda_1(\theta)\equiv-\lambda_0(2\pi-\theta)$ (indicated by the dashed line), resulting from the $\ZZ_2$ symmetry condition.
In such case, $\lambda_{\max}(0)$ is doubly degenerate, which contradicts the short-range correlation condition.

In order to circumvent such degeneracy, $\lambda_0(\theta)$ must flow along the real axis:
\begin{equation*}
\adjincludegraphics[scale=1,valign=c]{ 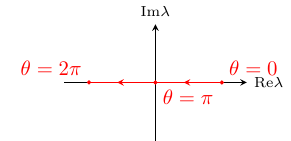}
\iffalse
    \begin{tikzpicture}[scale=0.8]
        \draw[->,>=stealth] (-55pt,0pt) -- (55pt,0pt)node[right] {\scriptsize $\rm{Re}\lambda$};
        \draw[->,>=stealth] (0pt,-35pt) -- (0pt,35pt)node[above] {\scriptsize  $\rm{Im}\lambda$}; 
        \draw[mid arrow,red] (40pt,0) --(0,0);
        \draw[mid arrow,red] (0,0)--(-40pt,0);
        \filldraw [red,fill=red] (40pt,0) circle (1pt) node[above right] {$\theta=0$};
        \filldraw [red,fill=red] (0,0) circle (1pt) node[below right] {$\theta=\pi$};
        \filldraw [red,fill=red] (-40pt,0) circle (1pt) node[above left] {$\theta=2\pi$};
    \end{tikzpicture}
    \fi
\end{equation*}
Consequently, $\lambda(\pi)=-\lambda(\pi)=0$, leading to divergence at $f(\theta=\pi)$.
In summary, $\lambda_{\max}(\theta)$ can never be a non-degenerate smooth function for half-integer spin chains. 

The flow of $\lambda_0(\theta)$ can take various scenarios, two of which are presented below:
\begin{enumerate}
    \item $\lambda_0(\theta)$ flows from $1$ to a real \emph{positive} number $\lambda_0(2\pi)$ along the real axis. 
        An inversion-related flow $\lambda_1(\theta)\equiv-\lambda_0(2\pi-\theta)$ emerges as a consequence of the symmetry condition in Eq.~\eqref{eq:transfer_mat_z2}:
       % \todo{Move Fig.~\ref{fig:specflow}(a) here. Delete Fig.~\ref{fig:specflow}(b)?}
       \begin{equation*}
\adjincludegraphics[scale=1,valign=c]{ 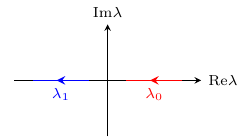}
\iffalse
    \begin{tikzpicture}[scale=0.9]
    \draw[->,>=stealth] (-50pt,0pt) -- (50pt,0pt)node[right] {\scriptsize $\rm{Re}\lambda$}; 
      \draw[->,>=stealth] (0pt,-30pt) -- (0pt,30pt)node[above] {\scriptsize $\rm{Im}\lambda$}; 
      \draw[mid arrow,red,thick](40pt,0)to node[below]{\scriptsize$\lambda_0$}(10pt,0);
      \draw[mid arrow,blue,thick](-10pt,0)to node[below]{\scriptsize$\lambda_1$}(-40pt,0);
  \end{tikzpicture}\fi
       \end{equation*}
        These two flows undergo a level crossing at $\theta=\pi$, leading to a cusp at $f(\pi)$.\label{item:positive}

    \item The flow of $\lambda_0(\theta)$ terminates at a \emph{negative} number $\lambda_0(2\pi)$.
        By further require $\lambda_0(\pi)$ to be non-vanishing, $\lambda_0(\theta)$ must flows initially along the real axis and then transitioning to the complex plane. 
        Due to symmetries, an additional branch $\lambda_1(\theta)$ emerges, which generally be different from $\lambda_0(\theta)$ along the real axis, yet forms a conjugate pair with $\lambda_0(\theta)$ upon entering the complex domain:
        \begin{equation*}
\adjincludegraphics[scale=1,valign=c]{ 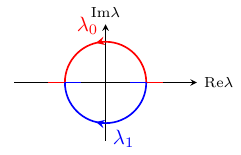}
        \iffalse
            \begin{tikzpicture}[scale=0.8]
                \draw[->,>=stealth] (-55pt,0pt) -- (55pt,0pt)node[right] {\scriptsize $\rm{Re}\lambda$};
                \draw[->,>=stealth] (0pt,-35pt) -- (0pt,35pt)node[above] {\scriptsize $\rm{Im}\lambda$}; 
                \draw[mid arrow,red,thick](24.5pt,0) arc (0:180: 24.5pt and 24.5pt);
                \draw[red,thick](24.5pt,0)--++(10pt,0);
                \draw[red,thick](-24.5pt,0)--++(-10pt,0);
                \draw[mid arrow,blue,thick](24.5pt,0) arc (0:-180: 24.5pt and 24.5pt);
                \draw[blue,thick](24.5pt,0)--++(-10pt,0);
                \draw[blue,thick](-24.5pt,0)--++(10pt,0);
                \node[above left,red] at (0,24.5pt) {$\lambda_0$};
                \node[below right,blue] at (0,-24.5pt) {$\lambda_1$};
            \end{tikzpicture}\fi
        \end{equation*}
        As depicted in the above figure, $\lambda_{0/1}(\theta)$ initially converge at a positive number $\lambda(\theta_0)$, split to conjugate pairs, encircle around origin, reconverge again at $\lambda(2\pi-\theta_0)\equiv-\lambda(\theta_0)$.
        Therefore, these flows lead to two cusps in $f(\theta)$ at $\theta_0$ and $2\pi-\theta_0$.\label{item:negative}
\end{enumerate}

\emph{Numerics. }
We now present our numerical results for a special class of open spin-$\frac{1}{2}$ chains, characterized by thermal states $\rho=Z^{-1} \ee^{-\beta H}$.
As we mentioned before, to demonstrate the anomaly phenomenon, the subgroup $U(1) \rtimes \ZZ_2 $ (generated by $\ee^{\ii \theta S^z_{\rm tot}}$ and $\ee^{\ii \pi S^x_{\rm tot}}$) of the full spin rotation is sufficient. 
Therefore, for the flexibility in numerics, we consider the nearest neighbouring $XXZ$ model 
\begin{align}
    \hat{H}=\sum_{j=1}^L \hat{S}_j^x\hat{S}_{j+1}^x + \hat{S}_j^y\hat{S}_{j+1}^y +\Delta\hat{S}_j^z\hat{S}_{j+1}^z
    \label{eq:xxz_ham}
\end{align}
%As $\hat{H}$ is real, $[\hat{T}(\theta)]^*=\hat{T}(-\theta)$. 
%Combining this observation with \eqref{eq:transfer_mat_z2}, we conclude that $\hat{T}(\theta)$ hosts an anti-unitary symmetry $W_x \KK$, where $\KK$ is complex conjugation.
%In the presence of such symmetry, eigenvalues of $\hat{T}(\theta)$ show up in complex conjugate pairs.
Numerical results for $\lambda_{0,1}(\theta)$ and $f(\theta)$ for spin-$\frac{1}{2}$ $XXZ$ chains are presented in Figure~\ref{fig:spinhalf}, with numerical details provided in the Supplemental Material~\cite{Sm}.

\begin{figure}[t]
    \centering
    \includegraphics[width=0.9\columnwidth]{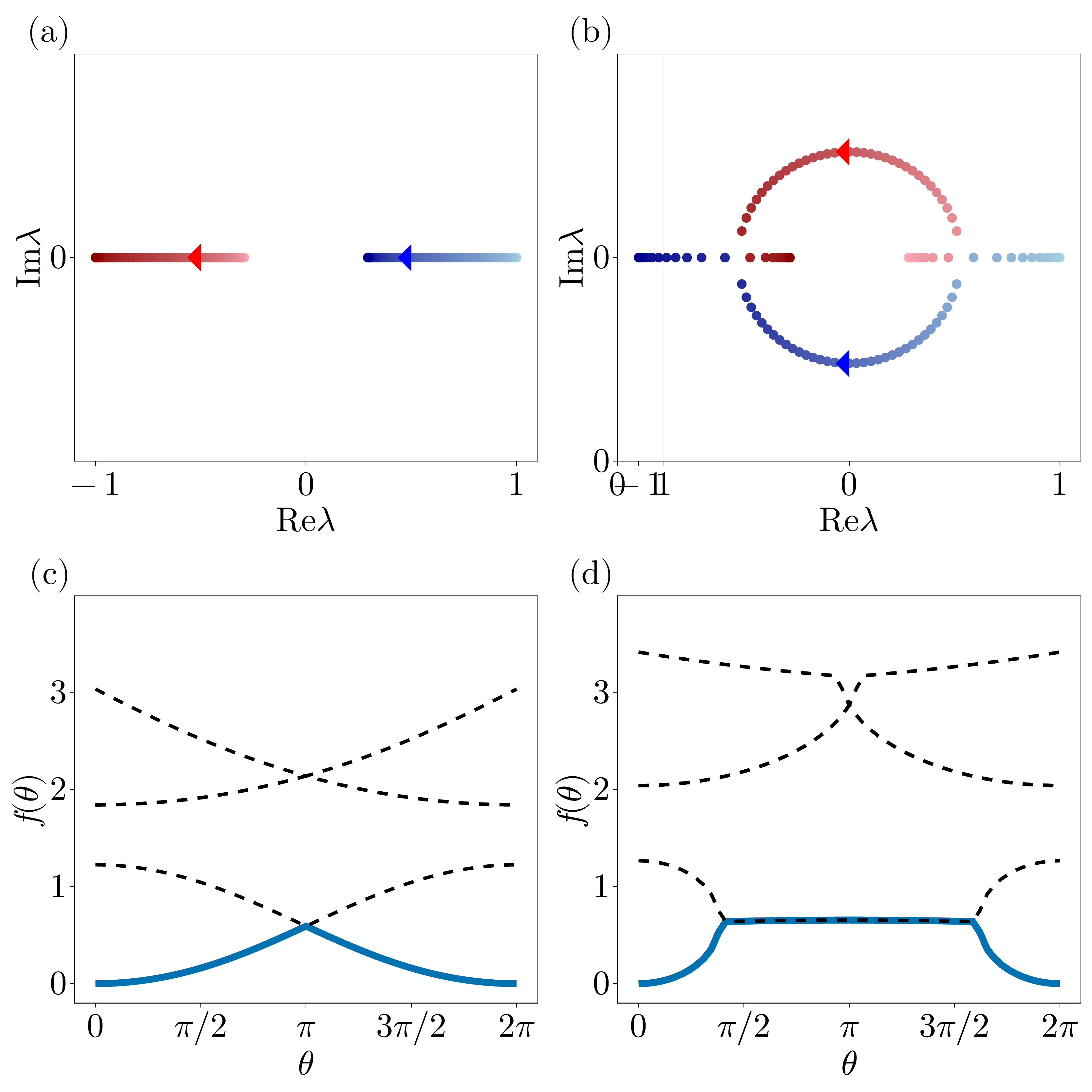}
    \caption{
    (a)/(b): Spectral flow of two leading eigenvalues of transfer matrix on the complex plane for a spin-1/2 chain \eqref{eq:xxz_ham} with $\Delta=+2/-2$.
    (c)/(d): The corresponding $f(\theta)$ (solid blue lines) as a function of $\theta$ for (a)/(b).
    Here we take six-step imaginary time evolution with $\delta\beta=0.1$.
    For comparison, we also plot logarithms of three more leading eigenvalues (dashed lines).}
        \label{fig:spinhalf}
    \end{figure}

When $\Delta=2$, the ground states of \eqref{eq:xxz_ham} belong to Ising AFM phases.
As illustrated in Figure~\ref{fig:spinhalf}(a), the two dominant eigenvalues of spatial transfer operator at finite $\beta$ are $\lambda_0(0)=1$ and $\lambda_1(0)=-1+\epsilon$, where the minus sign arises from translational symmetry breaking, and $\epsilon$ from the finite $\beta$ splitting~\cite{Sm}.
Therefore, it resembles flows in Item~\ref{item:positive}: with increasing $\theta$, $\lambda_0(\theta)$ and $\lambda_1(\theta)$ flows in the negative direction along the real axis, ending at $\lambda_0(2\pi)=1-\epsilon$ and $\lambda_1(2\pi)=-1$ respectively.
The dominant eigenvalue $\lambda_{\max}(\theta)$ transitions from $\lambda_0$ to $\lambda_1$ at $\theta=\pi$, resulting in a cusp of $f(\theta)$ at $\theta=\pi$, as depicted in Figure~\ref{fig:spinhalf}~(c). 

For the scenario with $\Delta=-2$, the ground states exhibit Ising ferromagnetic order. 
The two dominant eigenvalues of $\hat{T}$ at finite $\beta$ are $\lambda_0(0)=1$ and $\lambda_1(0)=1-\epsilon$, respectively~\cite{Sm}.
Therefore, similar as Item~\ref{item:negative}, $\lambda_0(\theta)$ flows from $1$ to $\lambda_0(2\pi)=-\lambda_1(0)=-1+\epsilon$, and $\lambda_1(\theta)$ flows from $1-\epsilon$ to $\lambda_1(2\pi)=-\lambda_0(0)=-1$.
As shown in Figure~\ref{fig:spinhalf}~(b), $\lambda_{0/1}(\theta)$ converge at $\pm\lambda(\theta_0)$, leading to two cusps in $f(\theta)$ at $\theta_0$ and $2\pi-\theta_0$, as illustrated in Figure~\ref{fig:spinhalf}(d). 

We also exam the behavior of $f(\theta)$ for spin-$1$ chains in Supplemental Material~\cite{Sm} and find the cusp could disappear (i.e. not topologically protected) as we tune $\beta$. 

\emph{Summary and outlook.} 
In this article, we proposed an open-closed correspondence that enable us to discuss concepts such as ``spectral flow'', ``flux threading'' for an open quantum system.
Using this correspondence, we propose a novel method to use the symmetry operator as a probe of the quantum anomaly in open systems.
We test this proposal by performing numerics in various thermal ensembles. 

In modern language, systems exhibiting quantum anomalies can find realization as reduced density matrix of symmetry-protected topological~(SPT) phases~\cite{Callan1985anomalies,Cheng2016}.
Within this paradigm, the quantity $\lrangle{\ee^{\ii\theta S^z_{\mathrm{tot}}}}$ corresponds to the topological  disorder operator~\cite{KadanoffCeva1971determination,Fradkin2017disorder,WuJianXu2021universal,WangChengMeng2021scaling} within SPT phases, a connection that warrants further investigation in future explorations~\cite{cai2024universal}. 
 
From the experimental perspective, our proposal to detect anomalies can be directly done by measure a single observable without changing the Hamiltonian or coupling the system to external fields.
An appealing platform where our approach could apply is the Rydberg atom arrays, where a programmable $XXZ$ Hamiltonian~\cite{Scholl2022xxz} could be realized.
In this platform, $\lrangle{\ee^{\ii\theta S^z_{\mathrm{tot}}}}$ could be obtained via post-processing data obtained from snapshot data, see details in Supplemental Material~\cite{Sm}.

\emph{Acknowledgement. } 
    We thank 
    Zhen Bi, 
    Meng Cheng, 
    Yichen Huang,
    Hui Zhai, and 
    Pengfei Zhang 
    for helpful discussions. 
    Z.Y.L. thank Ruizhen Huang and Guangyu Yu for useful discussions on numerics.
    This work is supported by MOST NO.~2022YFA1403902, CAS under contract No.~JZHKYPT-2021-08, NSFC Grant No.~12042505, NSFC Grant No.12342501.

    \emph{Note added.} 
    Upon finalizing this work, we notice a preprint by Hsin, Luo, and Sun \cite{hsin2023anomalies}, which also discusses the weak symmetry (called the average symmetry in the preprint).
    The anomaly phenomenon in their discussion is closer to \cite{ZhouLiZhaiLiGu2023}. 
    In addition, we are also aware of a few parallel works discussing disorder operators in various context  \cite{wang2023distinguishing,cai2024disorder,cai2024universal}

\bibliography{openanomaly}

\clearpage
\setcounter{equation}{0}
\renewcommand{\theequation}{S\arabic{equation}}

\setcounter{table}{0}
\renewcommand{\thetable}{S\arabic{table}}
\renewcommand{\theHtable}{\thetable}

\setcounter{figure}{0}
\renewcommand{\thefigure}{S\arabic{figure}}
\renewcommand{\theHfigure}{\thefigure}

\setcounter{section}{0}
\setcounter{secnumdepth}{3}

\title{Supplemental Materials: Detecting Quantum Anomalies in Open Systems}

\maketitle

\onecolumngrid

In this Supplemental Material, we provide a brief review of the flux-insertion argument for the LSM theorem~(Sec.~\ref{app:lsm}), discuss the relation between transfer matrix and correlators~(Sec.~\ref{app:src_density_matrix}), explore the behavior of $f(\theta)$ in thermal ensembles~(Sec.~\ref{app:thermal_ensemble}), present numerical details and further examples for $f(\theta)$~(Sec.~\ref{app:numerical_details}) and  experimental details~(Sec.~\ref{app:experimental_details}).

\section{Spectral flow and the LSM theorem}\label{app:lsm}
In this part, we review the flux-insertion argument for the LSM theorem\cite{LiebSchultzMattis1961}.
Instead of focusing on spin-$\frac{1}{2}$ chains, we examine a system consisting of $N$ hard-core bosons placed on a one-dimensional chain with $L$ sites, with onsite $U(1)$ charge conservation symmetry as well as translational symmetry $T_x$.
Such system is described by 
\begin{align}
    H=\sum_{i} (t\,b_i^\dagger b_{i+1}+h.c.) + \sum_{i}U n_i(n_i-1)+ \cdots
    \label{eq:hard_core_boson_ham}
\end{align}
where $\cdots$ denotes other symmetric terms.
Filling fraction $\nu \equiv \frac{N}{L} = \frac{q}{p}$ -- with $p$ and $q$ coprime numbers -- is imposed as an additional constraint on the whole Hilbert space.
The thermodynamic limit is achieved as both $N$ and $L$ approach infinity with $\nu$ fixed.
In the case where $\nu=\frac{1}{2}$, a $\mathbb{Z}_2$ particle-hole symmetry could be enforced, rendering the system equivalent to a spin-$\frac{1}{2}$ system with $U(1)\rtimes\mathbb{Z}_2$ symmetry, and Eq.~\eqref{eq:hard_core_boson_ham} could be transformed to spin-$\frac{1}{2}$ $XXZ$ model.

In such fractional filling system, the LSM theorem states that it must exhibit either gaplessness or ground state degeneracy in the thermodynamic limit. 
To illustrate this phenomenon, we introduce a magnetic flux parameterized by $\theta \in [0, 2\pi]$, achieved through the application of twisted boundary conditions, leading to 
\begin{align}
    H(\theta)=\sum_{i<L} (t\,b_i^\dagger b_{i+1}+h.c.) + (\ee^{\ii\theta}t\,b_L^\dg b_1+h.c.)+ \sum_{i}U n_i(n_i-1)+ \cdots
    \label{}
\end{align}
While $H(\theta)$ maintains the original $U(1)$ symmetry, its translational symmetry should be modified as 
\begin{align}
    T_x(\theta) \equiv \ee^{\ii\theta (b_1^\dagger b_1-\nu)} T_x : 
        b_i \to b_{i+1} \text{ for } i < L~;~
        b_{L} \to \ee^{-\ii\theta} b_{1}~.
    \label{eq:transl_theta}
\end{align}
It is easy to check that $[H(\theta),T_x(\theta)]=0$.
Note that phase factor $\ee^{-\ii\theta\nu}$ is included in the above definition, such that
\begin{align}
    [T_x(\theta)]^L = \ee^{\ii\theta(N-L\nu)} T_x^L = \hat{1}
    \label{eq:transl_theta_zl}
\end{align}
Consequently, the eigenvalues of $T_x(\theta)$, denoted as $\ee^{\ii k(\theta)}$, share the same \emph{discrete} set $\set{ \frac{2\pi n}{L}| n\in \ZZ}$ for any $\theta$.

We now explore the spectral flow of this system.
We denote the ground state of Eq.~\eqref{eq:hard_core_boson_ham} on a finite chain as $\ket{\psi_0}$ with energy $E_0$ and momentum $k_0$.
By varying $\theta$ from $0$ to $2\pi$, we trace the evolution of both the state, energy, and momentum, yielding $\ket{\psi_0(\theta)}$, $\lambda_0(\theta)$, and $k_0(\theta)$, respectively.
As $k_0(\theta)$ takes value in discrete set, we anticipate $k_0(\theta)=k_0$.
Combining with the observation that $T_x(2\pi) = \exp(-\ii 2\pi\nu) T_x$, we conclude that
\begin{align}
    T_x\ket{\psi_0(2\pi)} = \ee^{\ii (k_0+2\pi\nu)} \ket{\psi_0(2\pi)}
\end{align}
Hence, we identify $\ket{\psi_0(2\pi)}$ as a different eigenstate of Eq.~\eqref{eq:hard_core_boson_ham} from $\ket{\psi_0}$.
This procedure can be iterated, yielding at least $p$ distinct eigenstates, denoted as $\left\{ \ket{\psi_0(2\pi l)} \,\middle|\, l = 0, 1, \dots, p-1 \right\}$, where state $\ket{\psi_0(2\pi l)}$ carries momentum $k_0 + 2\pi l\nu$.
By imposing locality condition, one can show that energy gap between $\ket{\psi_0}$ and $\ket{\psi_0(2\pi l)}$ vanishes when $L\to\infty$, thus forbidding unique ground state in the thermodynamic limit.

\begin{figure}[htpb]
    \centering
    \includegraphics[scale=0.8]{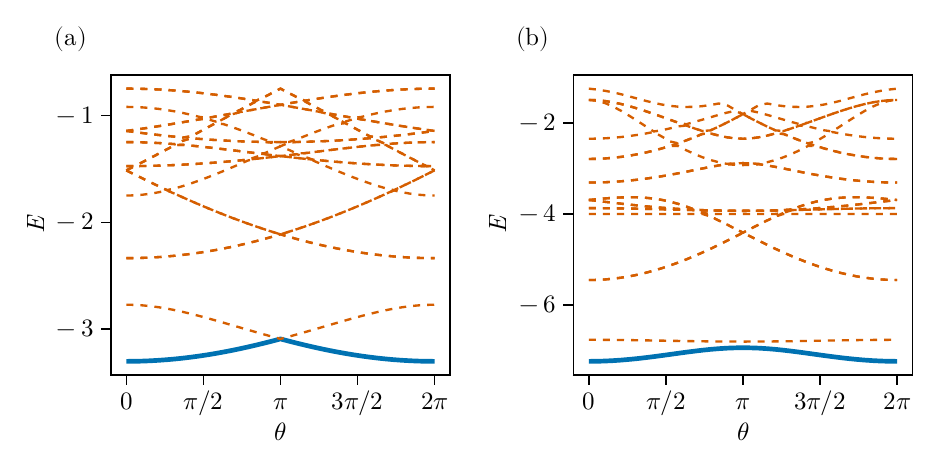}
    \caption{Spectral flow of (a) spin-$\frac{1}{2}$ $XXZ$ model~($L=6$, $\Delta=1.5$) and (b) spin-$1$ $XXZ$ model~($L=4$, $\Delta=1.5$).}
    \label{fig:energy_spectral_flow}
\end{figure}

Spectral flow of low-energy states for the spin-$\frac{1}{2}$ $XXZ$ chains and spin-$1$ $XXZ$ chains is calculated numerically, where the Hamiltonian with twisted boundary condition reads
\begin{align}
      \hat{H}(\theta)=\sum_{i=1}^{L-1} \frac{1}{2}(\hat{S}_i^+\hat{S}_{i+1}^{-} 
      + \hat{S}_i^- \hat{S}_{i+1}^+) +\Delta\hat{S}_i^z\hat{S}_{i+1}^z + \frac{1}{2}(\ee^{-\ii\theta}\hat{S}_L^+\hat{S}_{1}^{-} 
      + \ee^{\ii\theta}\hat{S}_L^- \hat{S}_{1}^+) +\Delta\hat{S}_L^z\hat{S}_{1}^z
        \label{}
\end{align}
As illustrated in Fig.~\ref{fig:energy_spectral_flow}, the spin-$\frac{1}{2}$ chain -- equivalent to $\nu=\frac{1}{2}$ hard-core boson system -- exhibits level crossing at $\theta=\pi$, while the ground state of spin-$1$ chain behaves as a smooth function.

\section{Transfer matrix and correlators in density matrix}\label{app:src_density_matrix}
In this part, we study the relation between correlators and spectrum of transfer matrix~(correlation spectrum) for translational invariant density matrix in an infinite 1D chain.
As in the main text, we consider the case where the density matrix can be well approximated as matrix product density operator~(MPDO)
\begin{align}
    \rho=\tr\left[ \cdots \hat{M}^{s_{j}s'_{j}}\hat{M}^{s_{j+1}s'_{j+1}}\cdots \right]\ket{\dots s_j s_{j+1},\dots}\bra{\dots s'_j s'_{j+1} \dots}
    \label{eq:mpdo}
\end{align}
Here, $\hat{M}^{ss'}=\sum_{\alpha,\alpha'=1}^{\chi}M^{ss'}_{\alpha\alpha'}\vket{\alpha}\vbra{\alpha'}$ with $\chi$ the internal bond dimension.
Transfer matrix $\hat{T}$ is defined as $\sum_s \hat{M}^{ss}$, and can be diagonalized as
\begin{align}
    \hat{T}=\sum_{n=0}^{\chi-1} \lambda_n\vket{r_n}\vbra{l_n}
    \label{eq:transfer_mat_eig_decomp}
\end{align}
where $\vket{r_n}$ and $\vbra{l_n}$ are the right and left eigenstates of $\hat{T}$ with (complex) eigenvalue $\lambda_n$, respectively. 
Normalization condition is imposed as $\left( l_m \middle| r_{n}\right)=\delta_{mn}$.
We adopt the convention that $\abs{\lambda_0} \geq \abs{\lambda_1} \geq \cdots$, where $\lambda_0$ are normalized to $1$.

$\lambda_n$'s form correlation spectrum, as they are related to correlators of local operators.
In the scenario where \emph{the dominant levels has no degeneracy}, i.e. $\abs{\lambda_{1}}<1$, the expectation value of a local operator $\hat{O}$ in the thermodynamic limit is given by
\begin{align}
    \lrangle{\hat{O}}\xrightarrow{L\to\infty}\vbra{l_0}\hat{T}({O})\vket{r_0}
    \label{eq:transfer_mat_expval}
\end{align}
where $\hat{T}({O})\equiv\sum_{\alpha\beta}\sum_{ss'}\left( M^{ss'}_{\alpha\beta}O_{s's}\right) \vket{\alpha}\vbra{\beta}$.
Two point correlator is expressed as 
\begin{align}
    \lrangle{\hat{O}^\dg_1\hat{O}_l}
    &\xrightarrow{L\to\infty}\sum_{n=0}^{\chi-1}\lambda_n^{l-2}\cdot \vbra{l_0}\hat{T}({O}^\dg)\vket{r_n}\cdot\vbra{l_n}\hat{T}({O})\vket{r_0}\notag\\
    &=\abs{\lrangle{\hat{O}}}^2+\sum_{n=1}^{\chi-1}\lambda_n^{l-2}\cdot \vbra{l_0}\hat{T}({O}^\dg)\vket{r_n}\cdot\vbra{l_n}\hat{T}({O})\vket{r_0}
    \label{eq:transfer_mat_corr}
\end{align}
So, when $l\gg1$,
\begin{align}
    \lrangle{\hat{O}^\dg_1\hat{O}_l}-\abs{\lrangle{\hat{O}}}^2\sim \ee^{-l/\xi}\,,~\text{with }\xi^{-1}=-\ln\abs{\lambda_a}
    \label{eq:transfer_mat_connected_corr}
\end{align}
where $a$ is the minimum number that satisfies $\vbra{l_a}\hat{T}(O)\vket{r_0}\neq0$.
In conclusion, when the dominant eigenvalues of correlation spectrum are non-degenerate, $\rho$ is short-range correlated. 

We now study local operator correlators for the case where modulus of dominant eigenvalues are degenerate.
For simplicity, we consider the two-fold (modulus) degeneracy case, and assume that $\lambda_1=1$. 
Generalization to cases where dominant eigenvalues have multiple degeneracy, and with different complex phases are straightforward.

For the current situation, the expectation value of local operator $\hat{O}$ on system in the thermodynamic limit is 
\begin{align}
    \lrangle{\hat{O}}\xrightarrow{L\to\infty}  \vbra{l_0}\hat{T}({O})\vket{r_0}+\vbra{l_1}\hat{T}({O})\vket{r_1}
    \label{eq:transfer_mat_deg_expval}
\end{align}
The two point correlator $\lrangle{\hat{O}_1\hat{O}_l}$ with $l\to \infty$ reads
\begin{align}
    \lrangle{\hat{O}^\dg_1\hat{O}_l} 
    \xrightarrow{L,l\to\infty}  \abs{\vbra{l_0}\hat{T}({O})\vket{r_0}}^2+\abs{\vbra{l_1}\hat{T}({O})\vket{r_1}}^2 +\vbra{l_0}\hat{T}({O}^\dg)\vket{r_1}\cdot\vbra{l_1}\hat{T}({O})\vket{r_0} +\vbra{l_1}\hat{T}({O}^\dg)\vket{r_0}\cdot\vbra{l_0}\hat{T}({O})\vket{r_1}
    \label{eq:transfer_mat_deg_corr}
\end{align}
Comparing Eq.~\eqref{eq:transfer_mat_deg_expval} and Eq.~\eqref{eq:transfer_mat_deg_corr}, we obtain the connected correlators
\begin{align}
    \lrangle{\hat{O}^\dg_1\hat{O}_l}_c
    \xrightarrow{L,l\to\infty}&
    \vbra{l_0}\hat{T}({O}^\dg)\vket{r_1}\cdot\vbra{l_1}\hat{T}({O})\vket{r_0}\notag
    +\vbra{l_1}\hat{T}({O}^\dg)\vket{r_0}\cdot\vbra{l_0}\hat{T}({O})\vket{r_1}\\
    &-\vbra{l_0}\hat{T}({O}^\dg)\vket{r_0}\cdot\vbra{l_1}\hat{T}({O})\vket{r_1}
    -\vbra{l_1}\hat{T}({O}^\dg)\vket{r_1}\cdot\vbra{l_0}\hat{T}({O})\vket{r_0}
    \label{}
\end{align}
which in general converge to a nonzero constant, distinct from the short-range correlated condition in Eq.~\eqref{eq:transfer_mat_connected_corr}.

In conclusion, density matrix $\rho$ is short-range correlated if and only if the dominant eigenvalue of transfer matrix is non-degenerate.

\section{$f(\theta)$ in thermal ensembles of spin chains}\label{app:thermal_ensemble}
In this section, we provide a detailed analysis for flow of $f(\theta)$ in thermal ensembles in a translational invariant $XXZ$ chain
\begin{align}
H=\sum_j \hat{S}_j^x\hat{S}_{j+1}^x + \hat{S}_j^y\hat{S}_{j+1}^y+\Delta\, \hat{S}_j^z \hat{S}_{j+1}^z + \cdots
    \label{eq:spin_ham}
\end{align}
Here, $\vec{\hat{S}}_j$ can be chosen as either spin-$\frac{1}{2}$ or spin-$1$.
Such system holds onsite symmetry $U(1)\rtimes\ZZ_2$, where $U(1)$ is spin rotation symmetry around $S_z$-axis, while $\ZZ_2$ is the $\pi$ rotation around axis in $S_x-S_y$ plane.

The thermal density matrix $\rho\equiv Z^{-1}\exp(-\beta H)$ satisfies the weak symmetry condition:
\begin{align}
    \mathbb{U}(g)\cdot\rho\cdot \mathbb{U}^\dg(g)=\rho\,,~\forall g\in U(1)\rtimes\ZZ_2
    \label{eq:density_mat_weak_sym}
\end{align}
We will focus on the case where $\rho$ is represented as MPDO as in Eq.~\eqref{eq:mpdo}.
From Eq.~\eqref{eq:density_mat_weak_sym}, symmetry constraints on local tensors is
\begin{align}
    M^{ss'}_{\alpha\beta}=[U(g)]_{st} [U^*(g)]_{s't'} [W(g)]_{\alpha\gamma} [W^*(g)]_{\beta\delta} M^{tt'}_{\gamma\delta} 
    \label{eq:local_tensor_sym}
\end{align}

We will analyze correlation spectrum of $\rho$ at zero temperature, generalize it to finite temperature, and finally study its flow under $U(1)$ symmetry action.

\subsection{Correlation spectrum of quantum phases at $0T$}
At zero temperature where $\beta\to \infty$, the system may exhibit different quantum phases depending on its microscopic details. 
For the symmetric phase with unique gapped ground state, all connected correlators decays exponentially as in  Eq.~\eqref{eq:transfer_mat_connected_corr}.
We mention that due to the LSM theorem, such phase only emerges in integer spin chains.
For gapless phase, the density matrix cannot be represented by MPDO, and is beyond our scope.
The system may also spontaneously breaks onsite or translational symmetries, exhibiting degeneracy in correlation spectrum, as we will discuss below.

For the case where internal $\ZZ_2$ symmetry is spontaneously broken, one may think that its local order parameter $\hat{S}^z$ acquires non-zero expectation value.
However, from symmetric condition in Eq.~\eqref{eq:density_mat_weak_sym}, the system is in a cat state with $\lrangle{\hat{S}^z}=0$.
The symmetry breaking phase is thus diagnosed by the non-vanishing correlator $\lrangle{{\hat{S}}^z_1{\hat{S}}^z_l}$ for $l\to\infty$.
Thus, Eq.~\eqref{eq:transfer_mat_connected_corr} no longer holds for $S_z$, and the dominant eigenvalues of $\hat{T}$ must be degenerate ($\abs{\lambda_1}=1$),
Furthermore, as $\vbra{l_0}T(S_z)\vket{r_0}=\vbra{l_1}T(S_z)\vket{r_1}=0$ due to symmetry, from Eq.~\eqref{eq:transfer_mat_deg_corr}, to get non-vanishing $S_z$ correlator, we require $\vbra{l_0}\hat{T}(S^z)\vket{r_1}\neq0$ as well as $\vbra{l_1}\hat{T}(S^z)\vket{r_0}\neq0$, meaning $\vket{r_0}/\vbra{l_0}$ carries opposite $\ZZ_2$ charge from $\vket{r_1}/\vbra{l_1}$ .
Let $\vket{r_n}/\vbra{l_n}$ pick up $(-1)^{s_n}$ under $\ZZ_2$ action, we conclude that $s_1=1-s_0$.

On the other hand, if the system spontaneously breaks translation symmetry, say with wave vector $k=\pi$, correlators of local operator should exhibit non-vanishing oscillation:
\begin{align}
    \lrangle{\hat{O}_1\hat{O}_l}-\abs{\lrangle{\hat{O}}}^2\sim (-1)^l\cdot C
    \label{eq:cdw_corr}
\end{align}
where $C$ is some nonzero constant.
Eq.~\eqref{eq:cdw_corr} is then incompatible with Eq.~\eqref{eq:transfer_mat_connected_corr}, which indicates that the dominant eigenvalues of $\hat{T}$ must be degenerate.
In fact, it is straightforward to check that to obtain Eq.~\eqref{eq:cdw_corr},  we require $\lambda_1=-\lambda_0=-1$.

According to the above discussion, we are able to determine properties of dominant eigenvalues of $\hat{T}$ for various gapped quantum phases, including the gapped symmetric phase,  the ferromagnetic phase along $z$ direction~($z$-FM phase), the valence bond solid~(VBS) phase, and the anti-ferromagnetic phase along $z$-direction~($z$-AFM phase), where results are summarized in Table~\ref{tab:dominant_level_quantum_phases}

\begin{table}[h!]
\centering
    \begin{tabular}{|c|c|c|c|c|}
    \hline
     & Symmetries & Correlators & Dominant levels of $\hat{T}$ \\
         \hline
       Gapped symmetric phase & preserving all symmetry & $\lrangle{\hat{O}_1^\dg \hat{O}_l}\sim \ee^{-l/\xi}$ & non-degenerate\\
       \hline
       $z$-FM phase & breaks $\ZZ_2$ symmetry & $\lrangle{\hat{S}^z_1\hat{S}^z_l}\neq 0$ for $l\to\infty$  & $\lambda_0=\lambda_1$, $s_0=1-s_1$ \\
       \hline
       VBS phase & breaks translation with $k=\pi$ & $\lrangle{\hat{O}_1^\dg \hat{O}_l}-\abs{\hat{O}}\sim (-1)^l\cdot C$ & $\lambda_0=-\lambda_1$, $s_0=s_1$\\
       \hline
       $z$-AFM  phase & breaks $\ZZ_2$ and translation with $k=\pi$ & $\lrangle{\hat{S}^z_1 \hat{S}^z_l}\sim (-1)^l\cdot C$ & $\lambda=-\lambda_1$, $s_0=1-s_1$\\
       \hline
    \end{tabular}
    \caption{Symmetries, characteristic behaviours of correlators, and degeneracy of dominant eigenvalues for various quantum phases on spin chains.
    Here, $\hat{O}$ denotes a generic local operator.}
    \label{tab:dominant_level_quantum_phases}
\end{table}

\subsection{Spectral flow at finite $\beta$}
At finite temperature with $\beta\gg0$, due to the Mermin-Wagner theorem, there will be no spontaneously symmetry breaking phase in a 1D chain.
Correlators of order parameters becomes exponential decay, whose correlation length $\xi\sim\beta$.
Thus, from Eq.~\eqref{eq:transfer_mat_connected_corr}, $-\ln\abs{\lambda_1}\sim\beta^{-1}$ and $\abs{\lambda_1}= 1-\epsilon(\beta)$ with $\epsilon(\beta)\sim\beta^{-1}$.

We now add $\ee^{\ii\theta S^z}$, and study spectrum of $\hat{T}(\theta)\equiv\sum_{ss'}\hat{M}^{ss'}\cdot \left( \exp[\ii\theta S^z] \right)_{s's}$.
$\ee^{\ii\pi S^x}$ is no longer a symmetry of $\hat{T}(\theta)$, instead,
\begin{align}
    \hat{T}(\theta)=W_x\cdot \hat{T}(-\theta)\cdot W_x^\dg
\end{align}
Note that $\hat{T}(\theta)=\pm\hat{T}(2\pi+\theta)$ for integer/half-integer spins.
In consequence, $\lambda(\theta)=\pm\lambda(2\pi-\theta)$ for integer/half-integer spin chains.
As argued in the main text, for density matrix of short-range correlation on a spin-$\frac{1}{2}$ chain, a flow starting from $\lambda_{0}$ would in general end at $-\lambda_1$ and vice versa.
While for a spin-$1$ chain, the starting and ending points of a flow could be the same one.

We mention that Hamiltonian in Eq.~\eqref{eq:spin_ham} hosts additional time reversal symmetry $\TT$, represented as $\exp[\ii \pi S^x_{\text{tot}}]\KK$.
Note that such symmetry commutes with $\exp[\ii\theta S^z]$, and thus still a symmetry for $\hat{T}(\theta)$.
In the presence of such anti-unitary symmetry, levels of $\hat{T}(\theta)$ must come in conjugate pairs:
\begin{align}
    \hat{T}(\theta)\vket{r}=\lambda\vket{r}
    ~\Longrightarrow~
    \hat{T}(\theta)\cdot\TT\vket{r}=\TT\cdot \hat{T}(\theta)\vket{r}=(\lambda)^*\cdot \TT\vket{r}
\end{align}
Combining with the above facts, we are able to sketch spectral flow for various quantum phases on spin-$\frac{1}{2}$ chains, and thus derive behavior of $f(\theta)$.

For the $z$-AFM or VBS phase, translational symmetry is broken at $0$T.
When temperature is turned on, $\lambda_1=-1+\epsilon(\beta)$.
Due to the $\TT$ symmetry, $\lambda_0(\theta)$ and $\lambda_1(\theta)$ both stick at the real axis, one from $1$ to $1-\epsilon(\beta)$, and another from $-1+\epsilon(\beta)$ to $-1$.
Consequently, the dominant level $\lambda_{\max}(\theta)$ transitions from $\lambda_0$ branch to $\lambda_1$ branch at $\theta=\pi$, resulting in a cusp of $f(\theta)$ at $\theta=\pi$, as verified by numerical results.

For the $z$-FM phase, at finite $\beta\gg0$, the degeneracy is split, where $\lambda_1=1-\epsilon(\beta)$ with $\epsilon(\beta)\sim\beta^{-1}$.
If $\lambda_0(\theta)$ flows from $\lambda_0=1$ to $-\lambda_1=-1+\epsilon(\beta)$ along the real axis, it would cross the origin point, leading to singularity.
To circumvent such an occurrence, and adhering to the constraints imposed by antiunitary symmetry, we propose the following spectral flow: $\lambda_{0/1}(\theta)$ intially converge at a positive number $\lambda(\theta_0)$, split to conjugate pairs due to the anti-unitary symmetry, encircles around origin, reconverge again at $-\lambda(\theta_0)$, and ultimately conclude at $-1$ and $-1+\epsilon$.
These behaviors of $\lambda_{0,1}(\theta)$ lead to two cusps in $f(\theta)$ at $\theta_0$ and $2\pi-\theta_0$, as shown in the main text.

\begin{figure}[t]
    \centering
   \includegraphics[scale=0.7]{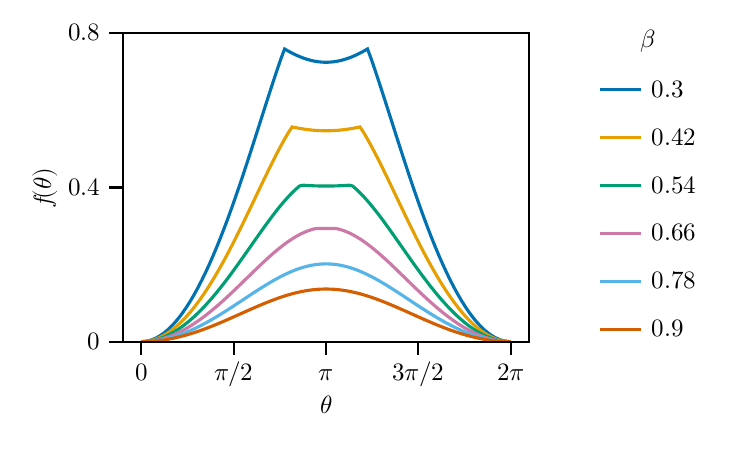}
    \caption{$f(\theta)$ as a function of $\theta$ for spin-1 model \eqref{eq:spin_ham} calculated in thermal ensembles at different temperature.
    We take $\Delta=2$ here.}
    \label{fig:spinone}
\end{figure}

We mention that for spin-1 chains, $f(\theta)$ could be smooth. 
To see this, we perform numerics to calculate $f(\theta)$ for the thermal ensemble described by the Hamiltonian in \eqref{eq:spin_ham} with $\Delta=2$ in various temperatures.
As illustrated in Figure~\ref{fig:spinone}, the cusps of $f(\theta)$ progressively converge, and eventually vanish as we increase $\beta$.

\section{More on Numerics}\label{app:numerical_details}
In this part, we will discuss numerical details for density matrix both in and out of thermal equilibrium.
\subsection{Thermal states}
\begin{figure}[htpb]
    \centering
    \includegraphics[scale=1]{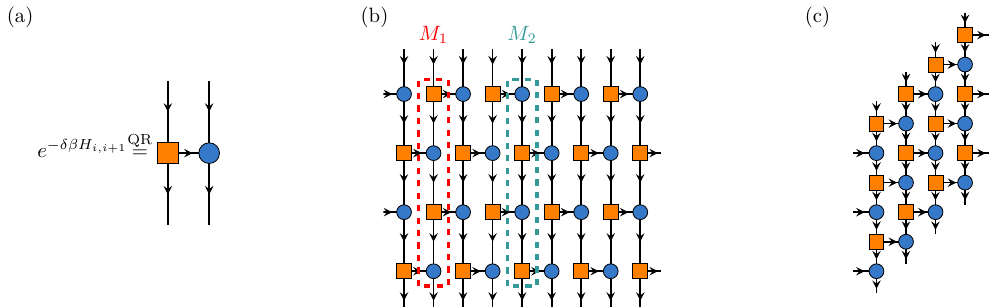}
    \caption{(a) QR decomposition of a Trotter gate. 
        (b) The conventional Suzuki-Trotter decomposition for a thermal density matrix by applying even and odd Trotter gates in even and odd rows respectively. 
        Such decomposition breaks translational symmetry, leading to two distinct transfer matrix $M_1$ and $M_2$.
(c) Thermal density matrix obtained by contracting $M_1$'s, which hosts translational symmetry.}
    \label{fig:tebd}
\end{figure}

Here, we present the numerical algorithm for calculation of correlation spectrum in thermal ensembles.
In particular, we employ the Suzuki-Trotter decomposition to approximate the thermal density matrix, representing $\rho$ as a 2D tensor network.
For Hamiltonian with nearest-neighboring interaction, we employ QR decomposition of $\ee^{-\delta\beta H_{i,i+1}}$, yielding two types of local tensors~(depicted as square and circle in Fig.~\ref{fig:tebd}(a)).
Both respect all internal symmetries.

In the conventional Suzuki-Trotter decomposition illustrated in Fig.~\ref{fig:tebd}(b),
\begin{align}
    \ee^{-\delta\beta H}=\ee^{-\delta\beta H_{\text{odd}}}\ee^{-\delta\beta H_{\text{even}}}+\mathcal{O}(\delta\beta^2)~,
    \label{}
\end{align}
and translation along spatial direction is explicitly broken due to the distinct operators $M_1$ and $M_2$ in even and odd columns.
Consequently, our spectral flow argument does not directly apply. 
To circumvent this subtlety, we explore an alternative decomposition, as depicted in Fig.~\ref{fig:tebd}(c), where only the contraction of $M_1$ operators takes place, thereby preserving translational symmetry.
Subsequently, we perform exact diagonalization for $T_1$ obtained by contracting vertical legs of $M_1$, whose eigenvalues are correlation spectrum. 
Similarly, spectrum of $\hat{T}(\theta)$ can be easily calculated numerically.
\subsection{Beyond thermal states}
\begin{figure}[t]
    \centering
   \includegraphics[scale=0.7]{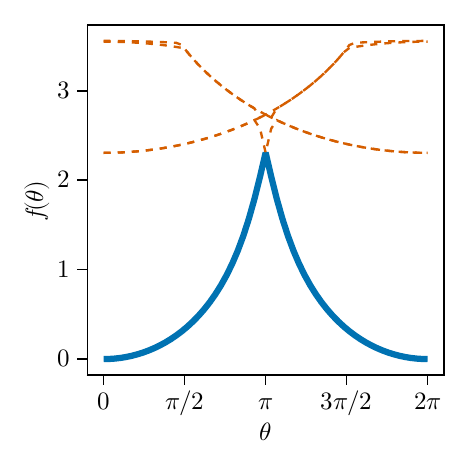}
    \caption{$f(\theta)$ as a function of $\theta$ for generalized thermal states.
    We take 2-step quantum operator with $\beta=0.1$, $H_1=H_{XXZ}(\Delta=2)$ and $H_2=H_{XXZ}(\Delta=-2)$}
    \label{fig:gene_therm}
\end{figure}

As mentioned in the main text, our detection scheme applies to density matrix beyond thermal ensembles, whose dynamics is governed by some quantum operations.
We here demonstrate this by considering the following density matrix
\begin{align}
    \rho = \cdots\ee^{-\beta_2 H_2}\ee^{-\beta_1 H_1}\ee^{-\beta_1 H_1} \ee^{-\beta_2 H_2}\cdots
\end{align}
where these Hamiltonians are imaginary-time dependent, and satisfy weak symmetry conditions. 
This state can be viewed as evolving from infinite temperature state, then acted on by quantum operator with one Kraus operator.
Numerical results for such states are presented in Fig.~\ref{fig:gene_therm}.

\section{Experimental details}\label{app:experimental_details}
In this part, we conduct a detailed analysis on experiment-related topics. 
We first present the concrete protocol on how to prepare and measure.
Various reasons which may affect the singularity are discussed.
\subsection{Preparation and measurement}
Our intended experimental platform is quantum simulators, such as Rydberg atom arrays. 
There are multiple methods for state preparation, where we list two here. 
First, we can create a symmetric pure state in a two-ladder system. 
The reduced density matrix of the upper half ladder satisfies the weak symmetry condition, and can be used to mimic the desired mixed state.
For the second scenario, we may create pure states with designed probability.
The resulting ensemble then mimic a mixed state.

For the measurement of $f(\theta)$, we note that it is diagonal in the $S^z$ basis, and is easy to measure in quantum simulators where single-shot measurements can be performed efficiently.
More specifically, we have
\begin{align}
  \lrangle{\ee^{\ii\theta S^z_{\mathrm{tot}}}}\approx\sum_{\{s_i^z\}}P(\{s_i^z\})\cdot \ee^{\ii\theta S^z_{\mathrm{tot}}}  
\end{align}
where $P(\{s_i^z\})=\bra{\{s_i^z\}}{\rho}\ket{\{s_i^z\}}$ the probability distribution.
Therefore, one can obtain snapshot data for each sample realization in $S^z$ basis, and then post-process the data to extract $f(\theta)$ without changing $\theta$.

\subsection{Error analysis}
\begin{figure}[t]
    \centering
   \includegraphics[scale=0.7]{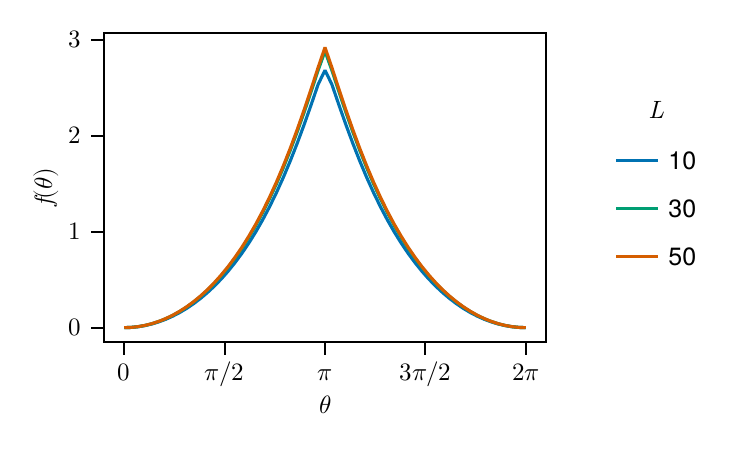}
    \caption{Open boundary condition with different lattice lengths. We take XXZ-model with parameters $\beta=0.1$ , $\Delta=2$.}
    \label{fig:finite_size}
\end{figure}

\emph{Finite size effect.}
In actual experimental settings, samples have finite size, and therefore the singularity of $f(\theta)$ will be smoothed out.
Despite this, for system sizes that are not too small, one can still see cusp(s) in $f(\theta)$, for both open and closed boundary conditions.
This singularity can be analyzed through finite-size scaling.
Fig.~\ref{fig:finite_size} illustrates the variation of $f(\theta)$ with respect to the system size $L$, revealing a progressively more discernible cusp-like behavior as $L$ increases.

\emph{Sampling.}
We note that the expectation value of $\lrangle{\exp(\ii S^z_{\rm tot})}$ decreases exponentially with system size.
Moreover, the number of measurements required to obtain reliable statistics does indeed increase exponentially as the system size grows.
Therefore, for both reasons, our protocol becomes impractical for very large system size. 
Combined with the previous discussion about finite size effect, our protocol is particularly well-suited for Noisy Intermediate-Scale Quantum (NISQ) devices~\cite{wei2023prx,cian2021many}.
    
\emph{Disorder.}    
While it is true that disorder breaks translation symmetry for one sample, the measurement of $f(\theta)$ in experiment involves a large ensemble of samples. 
The statistical averaging over many disordered samples effectively restores the translational symmetry.  
As a result, even in the presence of disorder, the singularity signal remains intact.

\end{document}